\documentclass[aps,prl,twocolumn,superscriptaddress,groupedaddress]{revtex4}
\usepackage{graphicx,amsmath,amssymb,appendix,afterpage,amsfonts,latexsym,color,dcolumn,bm,mathtools}
\pdfoutput=1
\newcommand{\beq}{\begin{equation}}
\newcommand{\eeq}{\end{equation}}
\newcommand{\bea}{\begin{eqnarray}}
\newcommand{\eea}{\end{eqnarray}}

\providecommand{\bra}[1]{\langle #1 \rvert}
\providecommand{\ket}[1]{\lvert #1 \rangle}

\newcommand{\ketbra}[2]{\left| {#1} \right\rangle\left\langle {#2}\right|}

\graphicspath{ {/} }

\usepackage{tikz}
\usepgflibrary{arrows}
\usetikzlibrary{decorations}
\usetikzlibrary{decorations.pathmorphing,patterns}
\usepackage{pgfplots}
\usetikzlibrary{scopes}
\usetikzlibrary{shapes, matrix}
\begin{document}

\title{Coherent Two-Dimensional Spectroscopy of a Fano Model} 

\author{Daniel Finkelstein-Shapiro}
\email[Corresponding author:$\;$]{daniel.finkelstein_shapiro@chemphys.lu.se}
\affiliation{Division of Chemical Physics, Lund University, Box 124, 221 00 Lund, Sweden}
\author{Felipe Poulsen}
\affiliation{Department of Chemistry, University of Copenhagen, DK 2100 Copenhagen, Denmark}
\author{T\~onu Pullerits}
\email[Corresponding author:$\;$]{tonu.pullerits@chemphys.lu.se}
\affiliation{Division of Chemical Physics, Lund University, Box 124, 221 00 Lund, Sweden}
\author{Thorsten Hansen}
\email[Corresponding author:$\;$]{thorsten@chem.ku.dk}
\affiliation{Department of Chemistry, University of Copenhagen, DK 2100 Copenhagen, Denmark}
\affiliation{Division of Chemical Physics, Lund University, Box 124, 221 00 Lund, Sweden}
\begin{abstract}
The Fano lineshape arises from the interference of two excitation pathways to reach a continuum. Its generality has resulted in a tremendous success in explaining the lineshapes of many one-dimensional spectroscopies - absorption, emission, scattering, conductance, photofragmentation - applied to very varied systems - atoms, molecules, semiconductors and metals. Unravelling a spectroscopy into a second dimension reveals the relationship between states in addition to decongesting the spectra. Femtosecond-resolved two-dimensional electronic spectroscopy (2DES) is a four-wave mixing technique that measures the time-evolution of the populations, and coherences of excited states. It has been applied extensively to the dynamics of photosynthetic units, and more recently to materials with extended band-structures. In this letter, we solve the full time-dependent third-order response, measured in 2DES, of a Fano model and give the new system parameters that become accessible. 
\end{abstract}

\maketitle

Whenever a discrete state and a continuum state interact and radiative transitions are allowed from the ground state to both, an interference occurs that gives rise to the Fano lineshape \cite{Fano1935,Fano1961,Rice1933,Beutler1935}. In the presence of dissipative processes, it has been shown that the generalized Fano equation consisting of a standard Fano plus a Lorentzian term is \cite{Fano1935,Fano1961,Friedrichs1948, Gallinet2011, Barnthaler2010, Finkelstein2015-1, Zhang2006}:
\begin{equation}
f(\epsilon,q,C)=\frac{(q+\epsilon)^2}{\epsilon^2+1} + \frac{C}{\epsilon^2+1}
\label{eq:Fano}
\end{equation}
where $q=\mu_e/n\pi V \mu_c$ and $\epsilon=(\omega_{L}-\omega_e)/\gamma_e$, $\mu_e$ is the transition dipole moment to the discrete excited state, $\mu_c$ the transition dipole moment to the continuum, $V$ the coupling of the discrete excited state to the continuum, $n$ the density of states of the continuum, $\hbar\omega_e$ is the energy of the discrete state relative to the ground state, $\omega_L$ is the radiation field frequency, and $\hbar\gamma_e=n\pi V^2$ is the linewidth of the excited state, induced by its coupling $nV^2$ to the continuum set of states. $C$ is the weight of the Lorentzian term and is related to the dissipative processes \cite{Zhang2006, Barnthaler2010, Gallinet2011, Finkelstein2015-1, Finkelstein2016}. 
\begin{figure}[ht]
\centering
\begin{tikzpicture}[scale=0.7]
\draw[thick] (0,0) -- (7cm,0) node[right]{$\ket{g}$};
\draw[thick] (0,3cm)--(4cm,3cm) node[right]{$\ket{e}$};
\draw[fill=gray] (5cm,1cm) rectangle (7cm,5cm) node[right]{$\ket{l}$} ;
\draw[->,thick] (2cm,0) --(2cm,2.9cm) node[midway, right] {$\mu_{e}$};
\draw [->,thick] (6.0cm,0cm)--(6.0cm,2.9cm)
node[midway, right] {$\mu_{c}$};
\draw[<->,thick] (3cm,3.1cm) to[out=45,in=135] node [sloped, above] {$V$} (6.0cm,3.1cm);
\draw[<-,thick,decorate,decoration=snake] (1.5cm,0) --(1.5cm,2.9cm) ;
\draw[<-,thick,decorate,decoration=snake] (5.5cm,0) --(5.5cm,3.25cm) ;
\end{tikzpicture}
\caption{\label{fig:FanoDissip} Energy levels and transitions of a Fano-type model (see main text for notations). Relaxation is allowed from the excited states to the ground state (marked as the squiggly transitions), as well as decoherence of all ground-excited coherences.}
\end{figure}
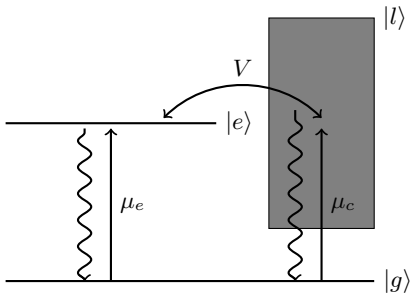
Since its original inception, the 1961 paper has resulted in more than 8,000 citations testifying to its ubiquity in physics, chemistry and materials science. The first Fano profiles were observed more than 80 years ago in atomic spectra, including photoionization of atoms and photodissociation of small molecules \cite{Beutler1935,Rice1933}. The field of study was established around scattering processes where the phenomena of coupling to a thermal bath resulting in dissipation or decoherence is not important, except in a few cases, notably pressure broadening \cite{Fano1963,Agarwal1982,Rzazewski1983}. 

\begin{figure}[h!]
\pgfmathdeclarefunction{gauss}{2}{%
 \pgfmathparse{exp(-((x-#1)^2)/(2*#2^2)}%
}
\pgfmathdeclarefunction{const}{2}{%
 \pgfmathparse{1}%
}
\begin{tikzpicture}
\begin{axis}[thick,
	no markers, domain=0:2, samples=30,
	axis lines=middle, xlabel=\empty, ylabel=\empty,
	every axis y label/.style={at=(current axis.above origin),anchor=south},
	every axis x label/.style={at=(current axis.right of origin),anchor=west},
	height=2.5cm, width=\columnwidth -1.5cm,
 	xtick=\empty, ytick=\empty,
	enlargelimits=false, clip=false, axis on top,
	grid = major
	]
	
	\addplot [ thick,red, domain=0.73:0.87] {gauss(0.8,0.02)};
	\addplot [ thick,black, domain=0.53:0.67] {gauss(0.6,0.02)};
	\addplot [ thick,black, domain=0.33:0.47] {gauss(0.4,0.02)};
	\addplot [ thick,black, domain=0.13:0.27] {gauss(0.2,0.02)};
	\node at (axis cs:0.2,1.25) {$k_1$};
	\node at (axis cs:0.4,1.25) {$k_2$};
	\node at (axis cs:0.6,1.25) {$k_3$};
	\node at (axis cs:0.8,1.25) {$k_S$};
	\draw [<->, thick, yshift=-0.3cm](axis cs:0.201,0) -- node[midway, below] {$\tau$} (axis cs:.399,0);
	\draw [<->, thick, yshift=-0.3cm](axis cs:0.401,0) -- node[midway, below] {$T$} (axis cs:.599,0);
	\draw [<->, thick, yshift=-0.3cm](axis cs:0.601,0) -- node[midway, below] {$t$} (axis cs:.799,0);
	\draw [->,thick](axis cs:0.92,0.4) -- node[midway, below, yshift = -0.13cm] {\scriptsize $-k_1\!+\!k_2\!+\!k_3$} (axis cs:1.02,0.1);
	\draw [->,thick](axis cs:0.92,0.6) -- node[midway, above, yshift = 0.13cm] {\scriptsize $k_1 \!- \!k_2 \!+\!k_3$} (axis cs:1.02,0.9);
	\draw [->,thick](axis cs:1.15,0) -- node[midway, below] {\small $\omega_\tau$} (axis cs:1.28,0);
	\draw [->,thick](axis cs:1.15,0) -- node[left, midway] {\small $\omega_t$} (axis cs:1.15,1);
	\draw [thin](axis cs:1.19,0.3) ellipse [x radius=0.15cm, y radius=0.15cm];
	\draw [thin](axis cs:1.19,0.3) ellipse [x radius=0.1cm, y radius=0.1cm];
	\draw [thin](axis cs:1.19,0.3) ellipse [x radius=0.05cm, y radius=0.05cm];
	
	\draw [thin](axis cs:1.24,0.7) ellipse [x radius=0.15cm, y radius=0.15cm];
	\draw [thin](axis cs:1.24,0.7) ellipse [x radius=0.1cm, y radius=0.1cm];
	\draw [thin](axis cs:1.24,0.7) ellipse [x radius=0.05cm, y radius=0.05cm];
	
	\draw [thin](axis cs:1.24,0.3) ellipse [x radius=0.1cm, y radius=0.1cm];
	\draw [thin](axis cs:1.24,0.3) ellipse [x radius=0.06cm, y radius=0.06cm];
	\draw [thin](axis cs:1.24,0.3) ellipse [x radius=0.025cm, y radius=0.025cm];
	
	\draw [thin](axis cs:1.19,0.7) ellipse [x radius=0.1cm, y radius=0.1cm];
	\draw [thin](axis cs:1.19,0.7) ellipse [x radius=0.06cm, y radius=0.06cm];
	\draw [thin](axis cs:1.19,0.7) ellipse [x radius=0.025cm, y radius=0.025cm];
\end{axis}
\end{tikzpicture}
\caption{In two-dimensional electronic spectroscopy, three pulses separated by time intervals $\tau$ and $T$ interact with a sample and an echo signal is generated at time $t$. The echo is emitted in a rephasing ($k_S=-k_1+k_2+k_3$) and a non-rephasing ($k_S=k_1-k_2+k_3$) direction. The signal is Fourier transformed with respect to the times $\tau$ and $t$ and presented as a 2-dimensional spectral map where the diagonal peaks correspond to populations, while off-diagonal peaks correspond to correlations between states.}
\label{fig:2D}
\end{figure}
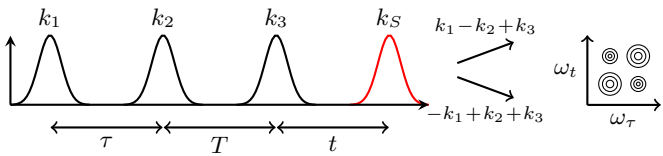

The synthesis and control of matter at the nanoscale opened up the observation of Fano profiles to dissipative systems \cite{Mirosh2010,Lucky2010,Lucky2012,Ojeda2009,Pakizeh2009} 
which in turn spurred a renewed interest in theoretical descriptions that would solve the problem in open quantum systems \cite{Robicheaux1995,Zhang2006,Kroner2008,Zhang2011,Finkelstein2015-1,Finkelstein2016}, and correctly account for the relaxation. Consequent with this desire to measure dissipation, more sophisticated spectroscopies are to be invoked. 
2D electronic spectroscopy (2DES) is arguably one of the best spectroscopies to measure decoherence and relaxation between states \cite{Gallagher1998,Hamm1998,Kjellberg2006,Cho2009}. It is a four-wave mixing technique that gives the full third-order response, from which one can infer the time evolution of the density matrix. 
Three time separated pulses with wave-vectors $k_1,k_2,k_3$ impinge on a sample and generate an echo in the $k_S=-k_1+k_2+k_3$ (rephasing) and $k_S=k_1-k_2+k_3$ (non-rephasing) directions (see Fig. \ref{fig:2D}). The signal is Fourier transformed along the $t$ and $\tau$ axes and we obtain 2D correlation maps for fixed time $T$, also called population time. 
It has been applied for over a decade to individual chromophores, photosynthetic proteins and even entire cells and has shed light on population transfer and vibronic coherence in natural systems \cite{Kjellberg2006,Zigmantas2006,Pullerits2013,Alster2014}. More recently, it has been used in structures with extended bands such as colloidal quantum dots, but the framework for simulating the 2D spectra for such systems is still incipient \cite{Wong2011,Seibt2013,Kobayashi2014}. 

Subsets of $\chi^{(3)}$ spectroscopies - focusing on only a portion of the third-order response - have been of interest to study Fano systems. 
Meier et al. analyzed theoretically the Fano signatures in a two-pulse four wave-mixing experiment in the case of single-and multiply excited discrete states \cite{Meier1995}. Nonlinear response of plasmon structures has been investigated numerically and experimentally \cite{Zhang2013,Metzger2014,Butet2014} where frequency tripling was found to be efficient in plasmonics dolmen supporting Fano resonances.
The current interest in mesoscopic devices that support Fano interferences as well as technological advances that permit the measurement of the full $\chi^{(3)}$ response with femtosecond resolution make the nonlinear response of a Fano model a timely topic. 

In this article, we present the analytical approach to the 2DES of a Fano system coupled to a Markovian bath which can be described by a Lindblad operator \cite{Lindblad1976,Gorini1976}. Both population relaxation and coherence dephasing are included in our model. 
Given the interest that exists around incorporating Fano systems into devices, spin filters and photonics, an understanding of the exact dephasing dynamics is essential. Moreover, a precise description of these processes can help us understand fundamentally the dynamics of excited states in band structures.

We start by writing the Fano Hamiltonian as (see Figure \ref{fig:FanoDissip}) \cite{Fano1961,Finkelstein2015-1}:
\begin{align}
&H=H_0+H_V+H_F \\
&H_0=E_{0}\ket{g}\bra{g}+E_e\ket{e}\bra{e}+\int dl \epsilon_l\ket{l}\bra{l} \nonumber \\
&H_V=\int dl \big[ V\ket{e}\bra{l}+V^*\ket{l}\bra{e} \big] \nonumber \\
&H_F=F \left[\mu_{e}\cos(\omega_L t)\ket{g}\bra{e}+\mu_{e}^*\cos(\omega_L t)\ket{e}\bra{g}\right] \nonumber \\
&+F \int dl \left[\mu_{c}\cos(\omega_L t)\ket{g}\bra{l}+\mu_{c}^*\cos(\omega_L t)\ket{l}\bra{g}\right],
\label{eq:Hamiltonian}
\end{align}
where $H_0$ is the site Hamiltonian, $H_V$ is the coupling of the discrete excited state to the continuum and $H_F$ is the interaction with the incident radiation field, allowing transitions from the ground state to the discrete excited state with transition dipole moment $\mu_e$ and to the continuum of states with transition dipole moment $\mu_c$. Without loss of generality we take $V,\mu_e,\mu_c$ to be real. 
We work under the typical wideband approximation which assumes all parameters related to the continuum are independent of the energy \cite{Fano1961}. However restrictive this condition might appear, it suffices for the parameters related to the continuum to be slowly varying within a linewidth of the discrete excited states and is well obeyed for all but a few exceptions. We also assume a linear dispersion such that $\omega_k=k/n$, $n$ being the density of states. This allows the expressions to be solved analytically, and although not all continua might be exactly described by this form, the resulting lineshape can describe even very complex systems succesfully (see for example \cite{Lombardi2010}). 

We now turn to an equivalent formulation of the problem by prediagonalizing the excited state and expressing the transition dipole moment in the new basis \cite{Rzazewski1983}. The new model consists of the ground state $\ket{g}$ and a new excited state manifold $\ket{k}$, which also includes the interaction with the discrete state. The new transition dipole moment from the ground state to a state $\ket{k}$ with energy $E(k)$ is given by \cite{Rzazewski1983}:
\begin{equation}
\mu_{kg}=\mu_0\frac{\epsilon+q}{\epsilon+i}
\label{Eq:effective_dipole}
\end{equation}
where $\epsilon=(E(k)/\hbar-\omega_e)/\gamma_e$ and $\hbar \gamma_e=n\pi V^2$. This formulation accounts for the discrete state by an energy-dependent transition dipole moment. 

\begin{figure}[ht]
\centering
\begin{tikzpicture}[scale=0.4]
\draw[thick] (2cm, 0cm) --(2cm, 1.5cm) ;
\draw[thick] (2cm, 1.5cm) --(2cm, 3cm) node[midway, left] {\footnotesize $\tau$};
\draw[thick] (2cm, 3cm) --(2cm, 4.5cm) node[midway, left] {\footnotesize $T$};
\draw[thick] (2cm, 4.5cm) --(2cm, 6cm) node[midway, left] {\footnotesize $t$};
\draw[->, thick] (2cm, 6cm) --(2cm, 7.5cm) ;

\draw[thick] (1.8cm, 1.5cm) --(2.2cm, 1.5cm) ;
\draw[thick] (1.8cm, 3cm) --(2.2cm, 3cm) ;
\draw[thick] (1.8cm, 4.5cm) --(2.2cm, 4.5cm) ;
\draw[thick] (1.8cm, 6cm) --(2.2cm, 6cm) ;

\draw[thick] (3.9cm, 0cm) --(3.9cm, 7.5cm);
\draw[thick] (3.9cm, 1.5cm) --(6.6 cm, 1.5cm) node[midway, yshift = 0.3cm] {\footnotesize $\ketbra{k}{g}$} node[midway, yshift = -0.3cm] {\footnotesize $\ketbra{g}{g}$};
\draw[thick] (3.9cm, 3cm) --(6.6 cm, 3cm) node[midway, yshift = 0.3cm] {\footnotesize $\ketbra{k}{k^\prime}$}; 
\draw[thick] (3.9cm, 4.5cm) --(6.6 cm, 4.5cm)node[midway, yshift = 0.3cm] {\footnotesize $\ketbra{k}{g}$};
\draw[thick] (3.9cm, 6cm) --(6.6 cm, 6cm) node[midway, yshift = 0.3cm] {\footnotesize $\ketbra{g}{g}$} node[midway, yshift =1cm] {\footnotesize $R_1$};
\draw[thick] (6.6 cm, 0cm) --(6.6 cm, 7.5cm) ;

\draw[->, thick] (3cm, 0.3cm) --(3.9cm, 1.5cm) node[xshift = -0.3cm, yshift = 0cm] {\footnotesize $k_1$};
\draw[->, thick] (7.5cm, 2cm) --(6.6cm, 3cm) node[xshift = 0.4cm, yshift = 0cm] {\footnotesize $-k_2$};
\draw[<-, thick] (7.5cm, 5.5cm) --(6.6cm, 4.5cm) node[xshift = 0.3cm, yshift = 0cm] {\footnotesize $k_3$};
\draw[<-, thick] (3cm, 7cm) --(3.9cm, 6cm) node[xshift = -0.3cm, yshift = 0cm] {\footnotesize $k_s$};

\draw[thick] (9.7cm, 0cm) --(9.7cm, 7.5cm);
\draw[thick] (9.7cm, 1.5cm) --(12.4cm, 1.5cm) node[midway, yshift = 0.3cm] {\footnotesize $\ketbra{g}{k^\prime}$} node[midway, yshift = -0.3cm] {\footnotesize $\ketbra{g}{g}$};
\draw[thick] (9.7cm, 3cm) --(12.4cm, 3cm) node[midway, yshift = 0.3cm] {\footnotesize $\ketbra{k}{k^\prime}$}; 
\draw[thick] (9.7cm, 4.5cm) --(12.4cm, 4.5cm)node[midway, yshift = 0.3cm] {\footnotesize $\ketbra{k}{g}$};
\draw[thick] (9.7cm, 6cm) --(12.4cm, 6cm) node[midway, yshift = 0.3cm] {\footnotesize $\ketbra{g}{g}$} node[midway, yshift = 1cm] {\footnotesize $R_2$};
\draw[thick] (12.4cm, 0cm) --(12.4cm, 7.5cm) ;

\draw[->, thick] (13.3cm, 0.3cm) --(12.4cm, 1.5cm) node[xshift = 0.4cm, yshift = 0cm] {\footnotesize $-k_1$};
\draw[->, thick] (8.8cm, 2cm) --(9.7cm, 3cm) node[xshift = -0.3cm, yshift = 0cm] {\footnotesize $k_2$};
\draw[<-, thick] (13.3cm, 5.5cm) --(12.4cm, 4.5cm) node[xshift = 0.3cm, yshift = 0cm] {\footnotesize $k_3$};
\draw[<-, thick] (8.8cm, 7cm) --(9.7cm, 6cm) node[xshift = -0.3cm, yshift = 0cm] {\footnotesize $k_s$};

\draw[thick] (2cm, -10.3cm) --(2cm, -9cm) ;
\draw[thick] (2cm, -9cm) --(2cm, -7.5cm) node[midway, left] {\footnotesize $\tau$};
\draw[thick] (2cm, -7.5cm) --(2cm, -6cm) node[midway, left] {\footnotesize $T$};
\draw[thick] (2cm, -6cm) --(2cm, -4.5cm) node[midway, left] {\footnotesize $t$};
\draw[->, thick] (2cm, -4.5cm) --(2cm, -3cm) ;

\draw[thick] (1.8cm, -9cm) --(2.2cm, -9cm) ;
\draw[thick] (1.8cm, -7.5cm) --(2.2cm, -7.5cm) ;
\draw[thick] (1.8cm, -6cm) --(2.2cm, -6cm) ;
\draw[thick] (1.8cm, -4.5cm) --(2.2cm, -4.5cm) ;

\draw[thick] (3.9cm, -10.3cm) --(3.9cm, -3cm);
\draw[thick] (3.9cm, -9cm) --(6.6 cm, -9cm) node[midway, yshift = 0.3cm] {\footnotesize $\ketbra{g}{k^\prime}$} node[midway, yshift = -0.3cm] {\footnotesize $\ketbra{g}{g}$};
\draw[thick] (3.9cm, -7.5cm) --(6.6 cm, -7.5cm) node[midway, yshift = 0.3cm] {\footnotesize $\ketbra{g}{g}$}; 
\draw[thick] (3.9cm, -6cm) --(6.6 cm, -6cm)node[midway, yshift = 0.3cm] {\footnotesize $\ketbra{k}{g}$};
\draw[thick] (3.9cm, -4.5cm) --(6.6 cm, -4.5cm) node[midway, yshift = 0.3cm] {\footnotesize $\ketbra{g}{g}$} node[midway, yshift = 1cm] {\footnotesize $R_3$};
\draw[thick] (6.6 cm, -10.3cm) --(6.6 cm, -3cm) ;

\draw[->, thick] (7.5cm, -10cm) --(6.6cm, -9cm) node[xshift = 0.4cm, yshift = 0cm] {\footnotesize $-k_1$};
\draw[->, thick] (6.6cm, -7.5cm) --(7.5cm, -6.6cm) node[xshift = 0cm, yshift = -0.3cm] {\footnotesize $k_2$};
\draw[->, thick] (3cm, -7cm) --(3.9cm, -6cm) node[xshift = -0.3cm, yshift = 0cm] {\footnotesize $k_3$};
\draw[<-, thick] (3cm, -3.5cm) --(3.9cm, -4.5cm) node[xshift = -0.3cm, yshift = 0cm] {\footnotesize $k_s$};
%
\draw[thick] (9.7cm, -10.3cm) --(9.7cm, -3cm);
\draw[thick] (9.7cm, -9cm) --(12.4cm, -9cm) node[midway, yshift = 0.3cm] {\footnotesize $\ketbra{k}{g}$} node[midway, yshift = -0.3cm] {\footnotesize $\ketbra{g}{g}$};
\draw[thick] (9.7cm, -7.5cm) --(12.4cm, -7.5cm) node[midway, yshift = 0.3cm] {\footnotesize $\ketbra{g}{g}$}; 
\draw[thick] (9.7cm, -6cm) --(12.4cm, -6cm)node[midway, yshift = 0.3cm] {\footnotesize $\ketbra{k^\prime}{g}$};
\draw[thick] (9.7cm, -4.5cm) --(12.4cm,-4.5cm) node[midway, yshift = 0.3cm] {\footnotesize $\ketbra{g}{g}$} node[midway, yshift = 1cm] {\footnotesize $R_4$};
\draw[thick] (12.4cm, -10.3cm) --(12.4cm, -3cm) ;

\draw[->, thick] (8.8cm, -10cm) --(9.7cm, 1-9.0cm) node[xshift = -0.3cm, yshift = 0cm] {\footnotesize $k_1$};
\draw[<-, thick] (8.8cm, -6.5cm) --(9.7cm, -7.5cm) node[xshift = -0.3cm, yshift = 0cm] {\footnotesize $-k_2$};
\draw[->, thick] (8.8cm, -7cm) --(9.7cm, -6cm) node[xshift = -0.3cm, yshift = 0cm] {\footnotesize $k_3$};
\draw[<-, thick] (8.8cm, -3.5cm) --(9.7cm, -4.5cm) node[xshift = -0.3cm, yshift = 0cm] {\footnotesize $k_s$};

\end{tikzpicture}
\caption{Double-sided Feynman diagrams used to calculate the third order response function. $R_1$ and $R_2$ are termed stimulated emission pathways (SE), and $R_3$ and $R_4$ are termed ground-state bleaching pathways. $R_2$ and $R_3$ are rephasing diagrams while $R_1$ and $R_4$ are non-rephasing diagrams}
\label{fig:Feynman-pathways}
\end{figure}
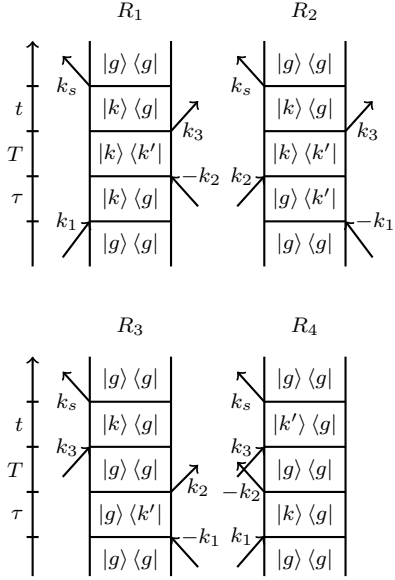

We calculate the 2DES response using perturbation theory \cite{Mukamel1995,Cho2009}. For a multiple level system, the response function consists of a number of pathways that grows combinatorially. Fortunately, in the limit of a continuum manifold the entire response function can be expressed as the integral of the Liouville pathways of an effective three-level system. We label any two states in the excited state manifold as $\ket{k}$ and $\ket{k'}$ and show their double-sided Feynman diagrams in Figure \ref{fig:Feynman-pathways}. The entire response function can be obtained by integrating over $k$ and $k'$. 

The experiment detects the electric field $E \propto -iP^{(3)}$.
The third-order polarization $P^{(3)}$ is written in terms of the response functions $R_i$ for the $i$th Feynman pathway \cite{Cho2009}. In the impulsive limit (where pulses have negligible duration):
\begin{equation}
\begin{split}
&P^{(3)}(t,T,\tau)=\left(\frac{i}{\hbar}\right)^3 \sum_{i=1}^4 R_i(t,T,\tau) e^{i\omega_L t \pm i\omega_L \tau}
\end{split}
\label{eq:general_equation}
\end{equation}
where the minus sign of the exponential is used for the rephasing pathways and the positive sign for the non-rephasing pathways and $\omega_L$ is the center frequency of the laser. The resulting expressions can be easily corrected for the finite bandwidth of the laser \cite{Nemeth2008}.
To illustrate how the calculation is done, we consider the ground-state bleach pathway of the rephasing signal ($R_3$). The calculations for the other pathways follow readily and are given in full detail in the Supplemental Information. Its expression is \cite{Mukamel1995,Cho2009}:
\begin{equation}
\begin{split}
R_3(t,T,\tau)&=\int dk \int dk' \lvert\mu_{kg}\rvert ^2 \lvert \mu_{k'g} \rvert ^2e^{-i\omega_{kg}t}e^{-i\omega_{gk'}\tau}\\
&\times F(t,T,\tau)
\end{split}
\end{equation}
where 
\begin{equation}
\begin{split}
F(t,T,\tau)&=\text{exp}[-g_{k'k'}^*(\tau)-g_{kk}^*(t)]
\end{split}
\end{equation}
where $g_{ba}(t)=\int_0^t d\tau_1 \int_0^{\tau_1} d\tau_2 C_{ba}(\tau_2)$ and $C_{ba}(t)=\frac{1}{\hbar^2}\langle U_{bg}(t)U_{ag}(0)\rho_B \rangle _B$. The potential $U_{ag}(t)=U_a(t)-U_g$ corresponds to the $a\leftrightarrow g$ transition, which fluctuates in time due to the interaction with the bath and $\langle \rangle_B$ is the average over the bath degrees of freedom. The lineshape functions $g(t)$ reveal information on the system-bath couplings. When the dynamics of the bath are much faster than those of the electronic system, we can invoke the Markovian approximation. In the case of a ground-excited coherence, the lineshape functions become \cite{Cho2009} $g_{aa}(t)=\gamma_{aa} t$. The Markovian approximation then corresponds to $C_{aa}(t)=\delta(t)\gamma_{aa}$: the bath looses its memory instantly. In this case, the rephasing pathway $R_3$ acquires a simple form:
\begin{equation}
\begin{split}
R_3(\tau,T,t)=&\int dk \int dk' \lvert\mu_{kg}\rvert ^2 \lvert \mu_{k'g} \rvert ^2 e^{-i\omega_{kg}t}e^{-i\omega_{gk'}\tau}\\
&\times e^{-\gamma_{kk}t-\gamma_{k'k'}\tau}
\end{split}
\end{equation}
The function $\gamma_{kk}$ can be thought of as the relaxation rate of the $gk$ coherence. Whenever the pathway evolves with an excited-excited coherence (such as $R_1$ and $R_2$), the dephasing rate is $\eta_{kk'}=(\gamma_{kk}+\gamma_{k'k'}-2\gamma_{kk'})$ where $\gamma_{kk'}$ is a correlation of the fluctuations of the states $k$ and $k'$. If the two states are completely uncorrelated, then we will have $\gamma_{kk'}=0$ and $\eta_{kk'}=\gamma_{kk}+\gamma_{k'k'}$. This is the case when the two energy levels lie on different coupled sites, for example an excitation on two isolated chromophores. If on the other hand the two energy levels lie on rigid sites, as is the case of a band structure, then we can expect for their fluctuations to be correlated and have $\eta_{kk'}\approx 0$.

We perform an extension of the wideband approximation for dissipative systems: the relaxation rates are slowly varying with respect to the energy of the continuum. To summarize, we use:
\begin{equation*}
\begin{split}
&\gamma_{kk}\to \gamma: \text{dephasing of the ground-continuum coherence} \\
& \gamma_{kk'} \to \gamma_{\text{corr}}: \text{correlation of two excited states of the} \\
&\text{continuum},0<\gamma_{\text{corr}}<\gamma \\
& \eta_{kk'} \to \eta: \text{dephasing of the coherence between two states} \\
&\text{in the continuum}
\end{split}
\end{equation*} 

$R_3$ becomes after we carry out the integrals in expression \eqref{eq:general_equation} and the Fourier transforms over $t$ and $\tau$:
\begin{equation}
\begin{split}
R_3(\omega_{\tau}, T, \omega_{t}) =& \mu_0^4\int dk\int dk^\prime \lvert\mu_{kg}\rvert ^2 \lvert \mu_{k'g} \rvert ^2 \\
&\times \frac{1}{(\omega_\tau-\omega_{k^\prime}-i\gamma)(\omega_t-\omega_{k}+i\gamma)}
\end{split}
\label{eq:general_equation_markovian}
\end{equation}
The closed form solution of the 2D signal corresponding to each of the Feynman pathways is \cite{SM} \footnote{corrected equation}: 
\begin{equation}
\begin{split}
&\text{Rephasing pathways}: \\
R_2(\epsilon_{\tau},T,\epsilon_t)&=(\mu_0^2n\pi)^2\frac{\Gamma^2(q^2+1)^2e^{-(2\gamma_e+\eta+1/T_{\text{pop}})T}}{(\epsilon_\tau - i)(\epsilon_t + i)} \\
R_3(\epsilon_{\tau},T,\epsilon_t)&=(\mu_0^2n\pi)^2h^*(\epsilon_t)h(\epsilon_\tau) \\
&\text{Non-rephasing pathways}: \\
R_1(\epsilon_{\tau},T,\epsilon_t)&=-(\mu_0^2n\pi)^2\frac{\Gamma^2(q^2+1)^2e^{-(2\gamma_e+\eta+1/T_{\text{pop}})T}}{(\epsilon_\tau + i)(\epsilon_t + i)}\\
R_4(\epsilon_{\tau},T,\epsilon_t)&=-(\mu_0^2n\pi)^2h^*(\epsilon_t)h^*(\epsilon_\tau) 
\end{split}
\label{eq:main-result}
\end{equation}
where $\epsilon_{\tau}=(\omega_{\tau}-\omega_e)/(\gamma_e+\gamma)$, $\epsilon_{t}=(\omega_{t}-\omega_e)/(\gamma_e+\gamma)$, $\Gamma=\gamma_e/(\gamma_e+\gamma)$. The $h$ function is defined as:
\begin{equation}
\begin{split}
\text{Re}(h(\epsilon))&= \Gamma\frac{\epsilon(q^2-1)-2q}{\epsilon^2+1}\\
\text{Im}(h(\epsilon))&= f(\epsilon,q_{\text{eff}},C)
\end{split}
\end{equation}
where $C=(1-\Gamma)(1+q^2\Gamma)$ and $q_{\text{eff}}=\Gamma q$ and $f(\epsilon,q,C)$ is defined in Eq. \eqref{eq:Fano}. The parameter $\Gamma$ is a measure of the contribution of dissipation: when $\Gamma=1$ no ground-excited dephasing is present and when $\Gamma\to 0$ the dephasing rate is much larger than $n\pi V^2/\hbar$. It is interesting to note that when $\Gamma\to 0 $, $q_{\text{eff}}\to 0$, that is the interference is suppressed and we obtain an anti-Lorentzian lineshape. The population relaxation to the ground state, assumed to be independent of the energy of the excited state is added as an additional decay rate $1/T_{\text{pop}}$ during the evolution populaton time $T$ for those pathways that evolve along the excited state manifold during $T$. These expressions constitute the main result of this work. In the limit of large $q$ these reduce to Lorentzian lineshapes as expected. 

\begin{figure}	
\includegraphics{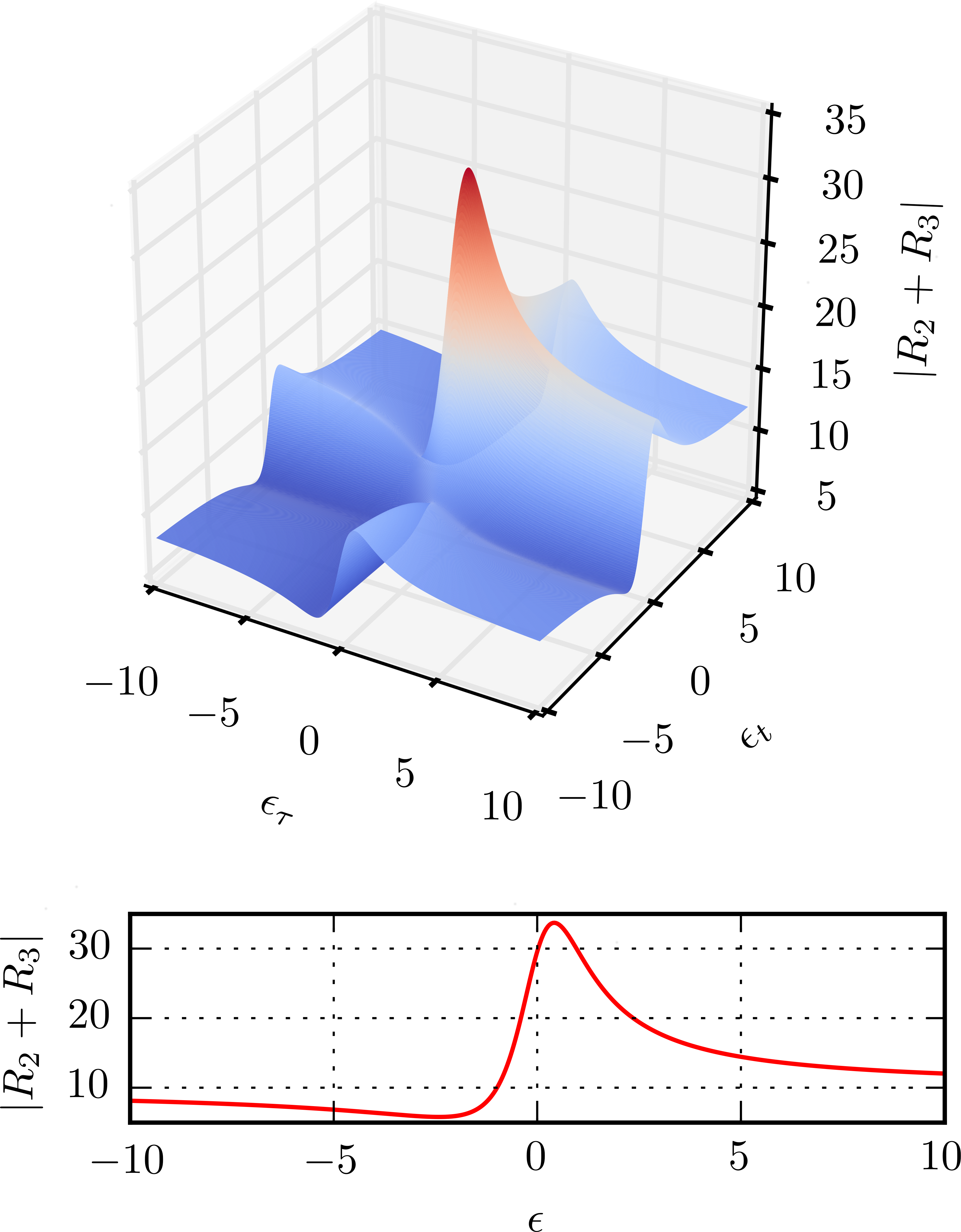}
\caption{Absolute value of the rephasing diagrams for $q=1$ with the diagonal plotted below.}
\label{fig:rephasing1}
\end{figure}

We plot in Figure \ref{fig:rephasing1} the absolute value of the rephasing pathways for $q=1$ where we see that the asymmetry of the one dimensional Fano lineshape is also present in the third order polarization signal. We can see from equations \eqref{eq:main-result} that the Fano profile comes from $R_3$ in the rephasing signal and $R_4$ for the non-rephasing signal. The diagonal signal of the rephasing spectrum is real and equal to (see Fig. \ref{fig:rephasing1}):
\begin{equation}
R_R^D(\epsilon)=(\mu_0^2n\pi)^2\bigg[\frac{\Gamma^2(q^2+1)^2e^{-(2
\gamma_e+ \eta+1/T_{\text{pop}})T}}{(\epsilon^2+1)} +|h(\epsilon)|^2 \bigg]
\end{equation}

\begin{figure}	
\includegraphics{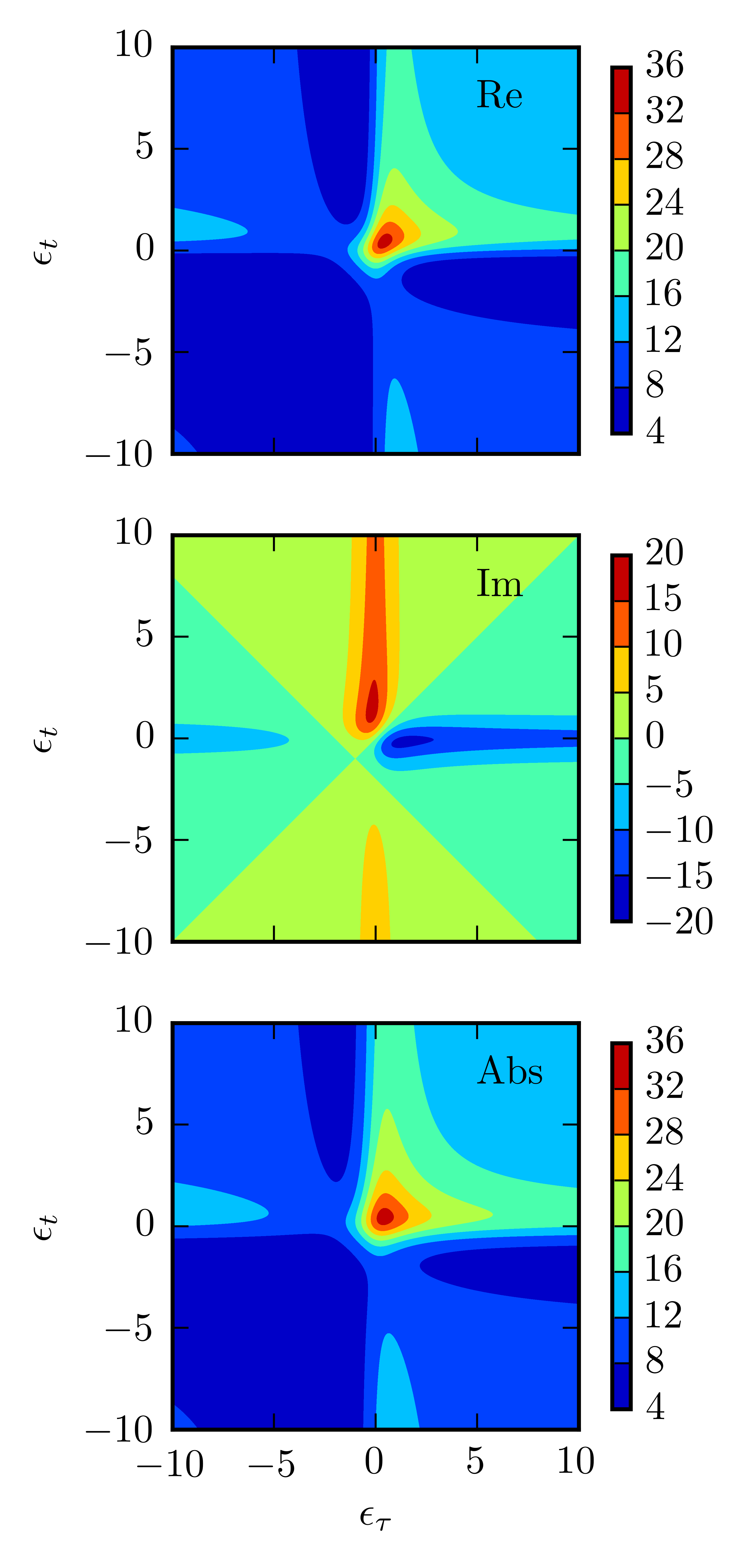}
\caption{Real, imaginary and absolute value of the Fano lineshape for $q=1$}
\label{fig:Fano2D-q1}
\end{figure}

The real, imaginary and absolute magnitude contour maps of the rephasing pathways for $q=1$ are shown in Figure \ref{fig:Fano2D-q1}. The asymmetry characteristic of Fano profiles can be discerned in the 2D plots by a breaking of the point symmetry about the origin. 
There are two types of information that we can extract from equations \eqref{eq:main-result}: a first family of parameters is obtained from fitting the spectra at each time slice $T$. These are a mix of structural and dynamical parameters $\Gamma$, $q$, $\gamma_e+\gamma$. The second family of parameters comes from fitting over several time slices where we obtain the dynamical functions $\eta=2(\gamma-\gamma_{\text{corr}})$. The fits provide structural information via the parameter $q$ and dynamical information by allowing us to extract the dephasings $\gamma_e$, $\gamma$. The strength of 2D spectroscopy is precisely to measure these dissipation constants (dephasing of coherences and relaxation of populations). Interestingly, it can also indicate whether the fluctuations on the excited state manifold are correlated (as measured by the correlation function $\gamma_{\text{corr}}$). \newline

We have presented the analytical solution to the full time-dependent third order polarization of a Fano model. Dephasings and correlations of the energy level fluctuations of the excited state are salient observables of the model. 
Originally confined to one or two laboratories, coherent two-dimensional spectroscopies are becoming more widespread and positioning themselves as the workhorse to study dynamical evolution of complex systems. They are possible in the IR region \cite{Hamm1998} and in the visible range. At visible wavelengths, they can be detected by photon echo \cite{Gallagher1998}, by fluorescence \cite{Tekavec2007,Karki2014}, by photocurrent \cite{Karki2014} or by photoelectrons \cite{Aeschlimann2011}. To each one of these methods corresponds a family of mesoscopic systems that may exhibit Fano profiles and that can be studied. The expressions derived in this article are general and can be tailored to each one of these cases and as such form the cornerstone of Fano studies in multidimensional spectroscopies. 
We expect that the advantages of decongesting the spectra that 2D affords will unveil new Fano interferences where they were masked before by spectral crowding or inhomogeneous broadening, and that the time resolution will provide a full characterization of the dissipative dynamics of coherences and populations to yield an unprecendented insight into the coupling of extended structures with the thermal bath. \newline

\textbf{Acknowledgements.} We gratefully acknowledge financial support from the Wenner-Gren foundation, the Lundbeck Foundation, the Swedish Research Council (VR) and the Knut and Alice Wallenberg Foundation. Collaboration within Nanolund is acknowledged. We wish to thank Sebastian R\"{o}ding and Julian Luettig from Prof. Brixner group for
bringing typos in Eq. 11 to our attention. \newline

\textbf{Accompanying software}. The scripts to generate the lineshapes described in this article can be downloaded from finoqs.wordpress.com. 

\bibliography{Fano2}

\end{document}